\documentclass[twocolumn,floats,prl,aps,showpacs]{revtex4}
\usepackage{graphicx}
\usepackage{amsfonts}
\usepackage{amsmath}
\usepackage{amssymb}
\usepackage{exscale}
\usepackage{color}
\usepackage{epsfig,psfrag,subfigure,subeqnarray}
\def\lan{\langle}
\def\ran{\rangle}

\def\dag{\dagger}

\def\vk{{\bf k}}

\def\v0{{\bf 0}}
\newcommand{\bd}{\begin{equation}}
\newcommand{\ed}{\end{equation}}
\newcommand{\be}{\begin{equation}}
\newcommand{\ee}{\end{equation}}
\newcommand{\bt}{\begin{split}}
\newcommand{\et}{\end{split}}

\newcommand{\bn}{\begin{align}}
\newcommand{\en}{\end{align}}
\newcommand{\bea}{\begin{eqnarray}}
\newcommand{\eea}{\end{eqnarray}}
\newcommand{\ba}{\begin{array}}
\newcommand{\ea}{\end{array}}
\newcommand{\nn}{\nonumber}

\begin{document}


\title{Scattering amplitudes for dark and bright excitons}
\author{Shiue-Yuan Shiau$^{1*}$\email[]{shiau.sean@gmail.com}, Monique Combescot$^2$, Roland Combescot$^3$, Fran\c{c}ois Dubin$^2$ and Yia-Chung Chang$^{4,1}$}\
\affiliation{(1) Department of Physics, National Cheng Kung University, Tainan, 701 Taiwan}
\affiliation{(2) Institut des NanoSciences de Paris, Universit\'e Pierre et Marie Curie,
CNRS, Tour 22, 2 Place Jussieu, 75005 Paris, France}
\affiliation{(3) Laboratoire de Physique Statistique, Ecole Normale Sup\'erieure, 24 rue Lhomond, 75005 Paris, France}
\affiliation{(4) Research Center for Applied Sciences, Academia Sinica, Taipei, 115 Taiwan}
\date{\today }

\begin{abstract}
Using the composite boson many-body formalism that takes single-exciton states rather than free carrier states as a basis, we derive the integral equation fulfilled by the exciton-exciton effective scattering from which the role of fermion exchanges can be unraveled. For excitons made of $(\pm1/2)$-spin electrons and $(\pm3/2)$-spin holes, as in GaAs heterostructures, one major result is that most spin configurations lead to brightness-conserving scatterings with equal amplitude $\Delta$, in spite of the fact that they involve different carrier exchanges. A brightness-changing channel also exists when two opposite-spin excitons scatter: dark excitons $(2,-2)$ can end either in the same dark states with an amplitude $\Delta_e$, or in opposite-spin bright states $(1,-1)$, with a different amplitude $\Delta_o$, the number of carrier exchanges being even or odd respectively. Another major result is that these amplitudes are linked by a striking relation, $\Delta_e+\Delta_o=\Delta$, which has decisive consequence for exciton Bose-Einstein condensation. Indeed, this relation leads to the conclusion that the exciton condensate can be optically observed through a bright part only when excitons have a large dipole, that is, when the electrons and holes are well separated in two adjacent layers. 
\end{abstract}


\maketitle
In contrast to the structureless $^4$He, a number of bosonic condensates have more than one component inherited from the internal spin and orbital degrees of freedom of their constituents, the superfluid then being multi-component.
Dipolar Bose gases \cite{stamper} and superfluid phases of $^3$He \cite{leggett} are prime examples.
This also occurs to excitons, which are composite bosons (cobosons for short)
made of one conduction electron and one valence hole. Since electrons and holes carry spins, so do excitons, their condensate depending on these internal degrees of freedom. Recently, it has been shown that signatures of exciton condensates were long held back by the missed fact that the lowest-energy states are dark \cite{cbc,cl}, that is, not coupled to light. This fact precludes a direct photoluminescence observation of exciton condensate in a very dilute regime. Unambiguous optical evidences for  Bose-Einstein condensation are bound to the density regime where dark and bright components coexist coherently \cite{cc}. The expected darkening\cite{Shilo_2013,alloing,Cohen_2016,beian,vortex} of the exciton gas upon cooling has been recently seen. Macroscopic spatial coherence of the bright component has also been revealed\cite{alloing,vortex}, in this way providing the most unambiguous evidence for the coexistence of dark and bright exciton condensates.

As a rule, the energy of a condensate depends weakly on its internal degrees of freedom, all possible ``spin" phases being essentially degenerate. These degeneracies are lifted once interactions are taken into account. One then has to determine which combination of the competing phases produces the lowest energy that rules the macroscopic properties of the condensate. In this Letter, we show that the relation between  brightness-conserving and brightness-changing scattering amplitudes constitutes a crucial element to characterize exciton Bose-Einstein condensates because the brightness-changing scattering provides the only channel to introduce a bright component into an otherwise dark condensate. The brightness and polarization of the exciton condensate are discussed in the light of this relation.

Having in mind GaAs heterostructures which have electrons with spin ($\pm1/2$) and holes with spin ($\pm3/2$), we first show that for all spin configurations in which the brightness is conserved, the scattering amplitudes have the same value, $\Delta$, despite differences in the carrier exchanges involved. By contrast, when two opposite-spin dark excitons, $(+2)$ and $(-2)$ scatter, they can either keep their darkness or become bright with spins $(+1)$ and $(-1)$, through different scattering amplitudes, $\Delta_e$ or $\Delta_o$, the number of carrier exchanges involved being even or odd respectively. We find that these scattering amplitudes are linked by $\Delta_e+\Delta_o=\Delta$, which is ultimately reasonable because the brightness-conserving scatterings contain even as well as odd numbers of carrier exchanges.

A salient consequence of the above relation between scattering amplitudes is that the lowest-energy state is optically dark in 3D and 2D systems. For these experimental topologies, the exciton condensate must be fully dark, thus ruling out any signature such as macroscopic spatial coherence through photoluminescence experiments. Yet, such a signature exists for bilayer heterostructures, with electrons and holes separately confined in two adjacent layers having a sufficiently large interlayer spatial separation \cite{cc}. The lowest-energy state then has coherent dark and bright components that allow a photoluminescence detection of the exciton condensate \cite{alloing,vortex}. 


\begin{figure}[t!]
\includegraphics[trim=1.5cm 4.3cm 0.8cm 3.5cm,clip,width=3.5in]{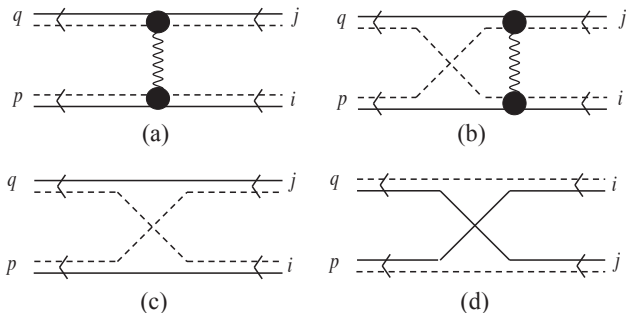}
 \caption{\small Shiva diagrams representing scatterings between two excitons:
  (a) direct-Coulomb scattering $\xi(_{p\,i}^{ q\,j})$; (b) hole-exchange-Coulomb scattering $ \xi_h^{in}(_{p\,i}^{ q\,j})$, the interaction taking place between the ``in" states $(i,j)$; (c) hole-exchange scattering $\lambda_h(_{p\,i}^{ q\,j})$; it is equal to the electron-exchange scattering $\lambda_e(_{p\,j}^{ q\,i})$, shown in (d). Electrons are represented by solid lines and holes by dashed lines. } 
  \label{Fig1}
 \end{figure}

Starting from 1970's, there has been a continuous focus on the study of scattering length for semiconductor excitons \cite{Keldysh1968,HaugPRB1975,ShumwayPRB2001,Noz,SCC2016}. However, the long-range character of Coulomb potential, and mostly the lack of appropriate procedure to handle carrier exchanges that occur along with repeated fermion-fermion interactions, have impeded significant progress. The realization, two decades ago, of Bose-Einstein condensates in ultracold atomic gases brought a new impetus, prompting scattering length studies for fermionic-atom dimers\cite{Pieri2000,Petrov2004,Brodsky2006,Birse2011,Alzetto2013}, and more recently for positronium atoms\cite{PlatzmannPRB1994,Ivanov2001,AdhikariPL2002,Chakraborty2004,Daily2015,Cassidy2005,Avetissian2014,Wang2014}. These studies commonly use fermions as elementary quantum objects. This includes the Pauli exclusion principle between the particle elementary constituents in the most  natural way. However, in doing 
 so, one loses the fact that, in a large sample volume, the two-pair scattered states stay very close to two non-interacting states. To take advantage of this physical fact, one has to represent the four fermions as two cobosons while properly handling fermion exchange between them. This precisely is what the coboson many-body formalism\cite{MoniqPhysreport,book,SCC2016} allows us to do. Here, we use this formalism to reveal, through Shiva diagrams, the fermion-exchange physics that occurs between bright and dark excitons and to grasp the topological equivalence of scattering processes that produces equal brightness-conserving scatterings. 

The coboson many-body formalism introduces

    \begin{figure}[t!]
\includegraphics[trim=1.5cm 7.5cm 0.8cm 6cm,clip,width=3.5in]{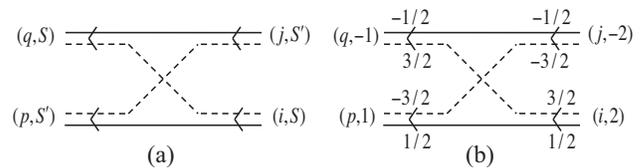}
 \caption{\small (a) Brightness-conserving scattering of two excitons $(S,S')$, with 
 $S{=}S'{=}(\eta2,\eta1)$ or 
 $(S{=}\eta2, S'{=}\eta'1)$ with
  $(\eta,\eta')={\pm}$. (b) Brightness-changing scattering induced by hole exchange turns two opposite-spin dark excitons into bright excitons. } 
  \label{Fig.2}
 \end{figure}

\noindent (i) the energy-like direct-Coulomb scattering, $\xi(_{p\,i}^{ q\,j})$, shown in Fig.~\ref{Fig1}(a), in which the exciton in state $i$ scatters to state $p$ while keeping its two fermions. The exciton index $i$ denotes its center-of-mass and relative-motion degrees of freedom. $\xi(_{p\,i}^{ q\,j})$ contains all possible fermion-fermion interactions between two excitons. 

\noindent (ii) the hole-exchange scattering, $\lambda_h(_{p\,i}^{ q\,j})$, shown in Fig.~\ref{Fig1}(c), in which the excitons $(i,j)$ exchange their holes, the excitons $p$ and $i$ having same electron. No fermion-fermion interaction occurs; so, this scattering is dimensionless. This hole-exchange scattering is topologically equivalent to an electron-exchange scattering within a $(i,j)$ permutation, $\lambda_h(_{p\,i}^{ q\,j})=\lambda_e(_{p\,j}^{ q\,i})$, as readily seen from their Shiva diagram representations, Figs.~\ref{Fig1}(c,d). 

\noindent (iii) the ``in" exchange-Coulomb scattering, defined as $\xi_h^{in}(_{p\,i}^{ q\,j})=\sum_{uv}\lambda_h(_{p\,u}^{ q\,v})\xi(_{u\,i}^{ v\,j})$, in which Coulomb interaction between the ``in" states $(i,j)$, is followed by a hole exchange (see Fig.~\ref{Fig1}(b)). A similar ``out" exchange-Coulomb scattering (not shown) has its Coulomb interaction between the ``out" states $(p,q)$.

In this formalism, it is easy to trace exciton spins using the Shiva diagrams of Fig.~\ref{Fig1}, each carrier keeping its spin when it scatters. Excitons made of $s$-spin electron and $m$-spin hole have a total spin $S=s+m$. For $s=\pm1/2$ and $m=\pm3/2$, the exciton spin can be $S=\pm1$ (excitons are bright) or $S=\pm2$ (they are dark, \textit{i.e}, not coupled to light). Three configurations can occur:

\noindent {\bf \textit{Case (1)}}: When two excitons have opposite-spin electrons and same-spin holes, $(1/2,-1/2;m,m)$, they form a dark and a bright exciton, ($S=2,S'=1)$ or ($S=-2,S'=-1$), depending on $m$. The spins, or brightness, of these $(S,S')$ excitons are conserved through direct-Coulomb scatterings  and hole-exchange scatterings. The same conclusion holds for $(s,s;3/2,-3/2)$ (see Fig.~\ref{Fig.2}(a)).

\noindent {\bf \textit{Case (2)}}: When two excitons have same-spin electrons and same-spin holes, $(s,s;m,m)$, they form two same-spin excitons, either dark $(S=\pm2)$, or bright $(S=\pm1)$, depending on $(s,m)$. The spin, or brightness, of these $(S,S)$ excitons is conserved through direct-Coulomb scatterings, electron-exchange scatterings and hole-exchange scatterings (see Fig.~\ref{Fig.2}(a)).

 \noindent {\bf \textit{Case (3)}}: When two excitons have opposite-spin electrons and opposite-spin holes, $(1/2,-1/2;3/2,-3/2)$, they can form either two opposite-spin dark excitons, $(2,-2)$, or two opposite-spin bright excitons, $(1,-1)$. Two opposite-spin dark excitons can keep their spins, or darkness, but they can also change into bright excitons through an odd number of hole, or electron, exchanges (see Fig.~\ref{Fig.2}(b)). Carrier exchange opens a brightness-changing channel which transforms two opposite-spin dark excitons into bright excitons, and vice versa. These dark-bright couplings split the dark-bright exciton quasi-degenerate subspace, making the present problem different from a standard elastic scattering problem. Case (3) thus requires a novel approach compared to the previous ones in which brightness is conserved.\

 \textbf{ \textit{Brightness-conserving scattering}}---The energy difference, $\Delta$, between the two-correlated-excitons ground state, $\mathcal{E}_2$, and two single-exciton ground states, $2\mathcal{E}_1$, scales as the inverse of the sample volume. It is commonly written in terms of the scattering length $a_s$. In 3D, dimensional arguments give this relation as
$\Delta\propto a_s/M L^3 $ (for $\hbar=1$), the prefactor\cite{Fetter} being equal to $4\pi$.

\begin{figure}[t!]
\includegraphics[trim=2cm 9.7cm 1.5cm 5.2cm,clip,width=3.5in]{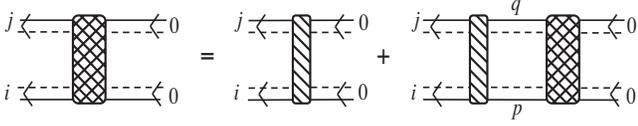}
 \caption{\small Effective scattering for exciton-exciton interaction, as given in Eq.~(\ref{laddereq}). } 
  \label{Fig.3}
 \end{figure}
\begin{figure}[t!]
\includegraphics[trim=1cm 4.3cm 1.7cm 5.2cm,clip,width=3.5in]{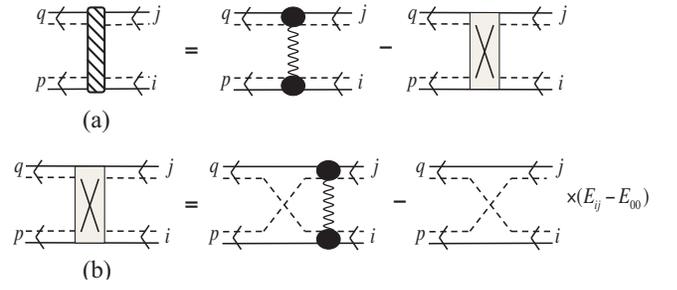}
 \caption{\small (a) Combination of elementary scatterings that rules the exciton-exciton effective scattering, as given in Eq.~(\ref{eq:zeta1}). (b) Exchange part, as given by the first expression of Eq.~(\ref{eq:exch}).}
  \label{Fig.4}
 \end{figure}

 
The coboson many-body formalism gives this energy difference as  \cite{SCC2016,S}
\be
\Delta=\mathcal{E}_2-2\mathcal{E}_1=\hat{\zeta}(_{0\,0}^{0\,0})\,,
\ee
with $\hat{\zeta}(_{p\,0}^{q\,0})$ solution of the integral equation shown in Fig.~\ref{Fig.3},
\be
\hat{\zeta}(_{\,p\,0}^{q\,0})=\zeta(_{\,p\,0}^{q\,0})+\sum_{ij\neq00}\zeta(_{p\,i}^{ q\,j})\frac{1}{E_{00}-E_{ij}}\hat{\zeta}(_{i\,0}^{i\,0})\, , \label{laddereq}
\ee
where $E_{ij}=E_i+E_j$ with $E_i$ being the $i$ exciton energy. The kernel scattering $\zeta(_{p\,i}^{ q\,j})$ has a direct part and an exchange part (see Fig.~\ref{Fig.4})
\begin{equation}
            \zeta(_{p\,i}^{ q\,j})
     =\xi(_{p\,i}^{ q\,j})-\xi^{exch}(_{p\,i}^{ q\,j})\, .\label{eq:zeta1}
        \end{equation}
 The exchange part, given by \cite{S}
        \bea
           \xi^{exch}(_{p\,i}^{ q\,j})
     =\xi^{in}(_{p\,i}^{ q\,j})+\lambda(_{p\,i}^{q \,j})\Big(E_{ij}-E_{00}\Big)\,\,\,\,\,\nonumber\\
     =\xi^{out}(_{p\,i}^{ q\,j})+\Big(E_{pq}-E_{00}\Big)\lambda(_{p\,i}^{ q\,j})\, ,
     \label{eq:exch}
   \eea
 contains the physically expected exchange-Coulomb scattering, $\xi^{in}(_{m\,i}^{ n\,j})$ or $\xi^{out}(_{m\,i}^{ n\,j})$, and a less obvious part constructed on the dimensionless carrier-exchange scattering, $\lambda(_{m\,i}^{ n\,j})$, multiplied by an energy difference that makes it energy-like and band-gap free, as physically required. Due to this $\lambda(_{m\,i}^{ n\,j})$ part, the kernel scattering, which is symmetrical with respect to ``in" and ``out"  states as seen by taking half the sum of the two expressions in Eq.~(\ref{eq:exch}), has the required time-reversal symmetry, $ \zeta(_{m\,i}^{ n\,j})= \big(\zeta(_{i\,m}^{ j\,n})\big)^*$. Also note that, without its $\lambda(_{m\,i}^{ n\,j})$ part, the scattering  would lead to a non-hermitian effective Hamiltonian\cite {Nous2016}.
 
   When the two excitons have same hole spin and opposite electron spins, $(1/2,-1/2;m,m)$, the exchange scattering that appears, $\xi_h^{exch}(_{p\,i}^{ q\,j})$, is constructed on $\lambda_h$, the two excitons exchanging their same-spin holes. And, similarly for $(s,s;3/2,-3/2)$ with $\xi_e^{exch}(_{p\,i}^{ q\,j})$ constructed on $\lambda_e$. For $(s,s;m,m)$, as for two same-spin dark or bright excitons, both electron exchange and hole exchange are possible, and $\xi_{eh}^{exch}(_{p\,i}^{ q\,j})$ is found equal to $\Big(\xi_{e}^{exch}(_{p\,i}^{ q\,j})+\xi_{h}^{exch}(_{p\,i}^{ q\,j})\Big)/2$.

   The first term in Eq.~(\ref{laddereq}) corresponds to the Born approximation, $\hat{\zeta}(_{0\,0}^{0\,0})\simeq{\zeta}(_{0\,0}^{0\,0})$. Since $\lambda_h(_{p\,i}^{ q\,j})$ and $\lambda_e(_{p\,i}^{ q\,j})$ are equal for $i=j$ or $p=q$, as readily seen from the Shiva diagrams of Fig.~\ref{Fig1}(c,d), we do have $\zeta_e(_{0\,0}^{0\,0})=\zeta_h(_{0\,0}^{0\,0})=\zeta_{eh}(_{0\,0}^{0\,0})$. So, exchanging a hole or an electron or both produces the same scattering amplitude at the Born level.\

 This result remains true at all orders in interaction. It is a direct consequence of
 $\xi(_{p\,i}^{ q\,j})=\xi(_{q\,j}^{ p\,i})$ and 
\bea
\label{sym11}
 \xi_h^{exch}(_{q\,j}^{ p\,i})=\xi_h^{exch}(_{p\,i}^{ q\,j})=\xi_e^{exch}(_{p\,j}^{ q\,i})=\xi_e^{exch}(_{q\,i}^{ p\,j})\, ,
 \eea
 which follows from the topological equivalence of their Shiva diagrams.
These relations readily give $\xi_h^{exch}(_{p\,0}^{\, q\,0})=\xi_e^{exch}(_{q\,0}^{\, p\,0})=\xi_e^{exch}(_{p\,0}^{\, q\,0})$. So, the first-order terms in Eq.~(\ref{laddereq}) are all the same whatever fermion exchanges are involved, $\zeta_h(_{p\,0}^{q\,0})=\zeta_e(_{p\,0}^{q\,0})=\zeta_{eh}(_{p\,0}^{q\,0})$. Now consider the second-order term in Eq.~(\ref{laddereq}). The one for two excitons having same hole spin can be rewritten as
 \bea
 \sum_{ij\neq00}\!\!\zeta_h(_{p\,i}^{ q\,j})\frac{1}{E_{00}{-}E_{ij}}\zeta_h(_{i\,0}^{j\,0})= \,\,\,\,\,\,\,\,\,\,\,\,\,\,\,\,\,\,\,\,\,\,\,\,\,\,\,\,\,\,\,\,\,\,\,\,\,\,\,\,\,\,\,\,\,\,\,\,\nonumber\\
\sum_{ij\neq00}\!\!\Big(  \xi(_{p\,i}^{ q\,j}) -\xi_h^{exch}(_{p\,i}^{ q\,j})\Big)\frac{1}{E_{00}{-}E_{ij}}\zeta_e(_{j\,0}^{i\,0})\, , \hspace{0.5cm}\label{sym}
 \eea
 in which we can replace $\xi_h^{exch}(_{p\,i}^{ q\,j})$ by $\xi_e^{exch}(_{p\,j}^{ q\,i})$. Interchanging the dummy indices $(i,j)$ in this exchange part then gives the above RHS as
   \begin{align}
 \sum_{ij\neq00}\zeta_e(_{p\,i}^{ q\,j})\frac{1}{E_{00}-E_{ij}}\zeta_e(_{i\,0}^{j\,0})\, ,
 \end{align}
 with $\zeta_e$ possibly replaced by $\zeta_{eh}$ through a similar procedure. By iterating the argument, we end up with $\hat{\zeta}_h(_{p\,0}^{q\,0})=\hat{\zeta}_e(_{p\,0}^{q\,0})=\hat{\zeta}_{eh}(_{p\,0}^{q\,0})$. The key for this surprising equivalence is that these scatterings start with two same-state excitons $(0,0)$.

\noindent \textbf{\textit{Brightness-changing scatterings}}---In Case (3), the brightness-changing channel couples the two-dark-exciton state $(2,-2)$ to the two-bright-exciton state $(1,-1)$. In calculating their scatterings, one can neglect the very small interband-Coulomb processes that produce the dark-bright energy splitting, and take these exciton states as degenerate. This degeneracy is lifted by interactions. As shown in the Appendix, the components of the bright state, $b_{00}$, and of the dark state, $d_{00}$, in the resulting eigenstates fulfill
\be
0=(b_{00}\pm  d_{00})\Big(\hat{\zeta}_{eh}^{\pm}(_{0\,0}^{ 0\,0})- \Delta^{\pm}\Big)\, ,\label{Deltac+d}
\ee 
where $\hat{\zeta}_{eh}^{\pm}$ is obtained from Eq.~(\ref{laddereq}) with $\zeta$ replaced by $ \zeta_{eh}^{\pm}=\xi\mp\xi_{eh}^{exch}$, and $\Delta^{\pm}$ are the eigen-energy differences. The above equation either gives $\mathcal{E}_2=2\mathcal{E}_1+\hat{\zeta}_{eh}^{+}(_{0\,0}^{ 0\,0})$ and $d_{00}=b_{00}$, or $\mathcal{E}_2=2\mathcal{E}_1+\hat{\zeta}_{eh}^{-}(_{0\,0}^{0\,0})$ and $d_{00}=-b_{00}$. 
From this solution, we can deduce the effective scatterings between two dark excitons ($\Delta_{dd}$), between two bright excitons ($\Delta_{bb}$) and between a dark and a bright exciton,
 $(\Delta_{db},\Delta_{bd})$, defined through
\be
\begin{bmatrix}
\Delta_{bb}-\hat{\zeta}_{eh}^{\pm}& \Delta_{bd}\\
\Delta_{db}& \Delta_{dd}-\hat{\zeta}_{eh}^{\pm}
\end{bmatrix}\begin{bmatrix}b_{00}\\ \pm b_{00}\end{bmatrix}=0\,.
\ee
This yields
\begin{subeqnarray} \label{10}
\frac{1}{2}\Big(\hat{\zeta}_{eh}^{+}(_{0\,0}^{0\,0})+\hat{\zeta}_{eh}^{-}(_{0\,0}^{0\,0})\Big)=\Delta_{dd}=\Delta_{bb} \equiv \Delta_e\,,\\
\frac{1}{2}\Big(\hat{\zeta}_{eh}^{+}(_{0\,0}^{0\,0})-\hat{\zeta}_{eh}^{-}(_{0\,0}^{0\,0})\Big)=\Delta_{db}=\Delta_{bd}\equiv\Delta_o\, .
\end{subeqnarray}
By noting that $\Delta=\hat{\zeta}_{eh}^{+}(_{0\,0}^{0\,0})$, we readily find that these scattering amplitudes are related by
\be\label{eq11}
\Delta=\Delta_e+\Delta_o\,.
\ee

At the Born level\cite{OBM,book}, the brightness-conserving scattering $\Delta_e$ reduces to the direct-Coulomb scattering $\xi(_{0\,0}^{0\,0})$, which is equal to zero since $\xi(_{i\, i}^{nj})=0$. The brightness-changing scattering reduces to $\Delta_o \simeq -\xi^{in}(_{0\,0}^{0\,0})
 =\xi_D(a_X/L)^DR_X$ where $a_X$ is the 3D exciton Bohr radius, $R_X=1/2\mu a_X^2$ the exciton Rydberg and $D$ the space dimension, with $\xi_3=26\pi/3$ as first obtained by Keldysh-Kozlov\cite{Keldysh1968}, and $\xi_2=8\pi-315 \pi^3/512$ in 2D for electrons and holes in the same quantum well\cite{MA}, this value turning negative when the carriers are in two distant planes\cite{CCAD}.

 The integral equation (\ref{laddereq}) allows going beyond the Born approximation. From Eq.~(\ref{10}), we see that $\Delta_o$ contains an odd number of fermion exchanges while in $\Delta_e$, this number is even, as physically expected because dark and bright states are coupled by carrier exchange while two exchanges reduce to an identity. The calculation of $\hat{\zeta}(_{0\,0}^{0\,0})$ requires the knowledge of the kernel scattering $\zeta(_{i\, p}^{ j\, q})$ which is given in terms of $(\lambda,\xi,\xi^{in})$, all of which depend on single-exciton wave functions\cite{MoniqPhysreport,SYAnnals}. It has been shown that, in 3D for equal carrier masses \cite{ShumwayPRB2001}, $\hat{\zeta}_{eh}^{-}\simeq3\hat{\zeta}_{eh}^{+} \simeq9 \pi (a_X/L)^3 R_X$, these scatterings being rather insensitive to mass ratios up to $m_h/m_e\simeq10$. As $\hat{\zeta}_{eh}^{+}=\Delta$, this gives $\Delta_e\simeq-\Delta$. By contrast, in biased heterostructures with well-separated electrons and holes, we expect smaller exchange contributions; so, $\hat{\zeta}_{eh}^{-}\simeq\hat{\zeta}_{eh}^{+}$ which will make $\Delta_o$ smaller than $\Delta$. The numerical resolution of the integral equations that give these effective scatterings in the physically relevant configurations will be addressed in a near future.

\textbf{\textit{ Condensate brightness and polarization}}---The above scattering amplitudes between dark $S=\pm2$ and bright $S=\pm1$ excitons lead to an effective Hamiltonian 
\bea \label{12}
H_{eff}=\varepsilon_{bd}\!\!\sum_{S=\pm1}B^\dag_{S}B_{S}{+} \frac{\Delta}{2}\!\!\sum_{S=(\pm2,\pm1)}\!\! B^\dag_{S}B^\dag_{S}B_{S}B_{S}  \hspace{10mm}   \nonumber\\
{+}\Delta\!\!\sum_{S=\pm2}\sum_{S'=\pm1}\!\! B^\dag_{S}B^\dag_{S'}B_{S'}B_{S}{+}\Delta_e\!\! \sum_{S=(2,1)}\!\! B^\dag_{S}B^\dag_{-S}B_{-S}B_{S} \nonumber \\
{+}\Delta_o \Big(  B^\dag_{2}B^\dag_{-2}B_{-1}B_{1} {+}\textit{h.c.}  \Big)\, ,  \hspace{30mm}
\eea
where $B^\dag_{S}$ is the $S$-spin exciton creation operator and $\varepsilon_{bd}$ is the bright-dark splitting, the dark exciton energy being taken as zero. The relation (\ref{eq11}) between scattering amplitudes gives $H_{eff}$ as $H'_{eff}{+}\widehat{N}(\widehat{N}{-}1)\Delta/2$ where $\widehat{N}{=}\sum_S B^\dag_{S}B_{S}$ is the total exciton-number operator and
\bea \label{13}
H'_{eff}=\varepsilon_{bd}\sum_{S=\pm1}B^\dag_{S}B_{S}                 \,\,\,\,\,\,\,\,\,\,\,\,\,\,\,\,\,\,\,\,\,\,\,\,\,\,\,\,\,\,\,\,\,\,\,\,\,\,\,\,\,\,\,\,\,\,\,\,\,\,\,\,\,\,\,\,\,\,\,\,\,\,\,              \nonumber\\ 
{-}\Delta_o  \Big( B^\dag_{2}B^\dag_{-2}{-}B^\dag_{-1}B^\dag_{1} \Big)
 \Big( B_{2}B_{-2}{-}B_{-1}B_{1} \Big)\,.                            
\eea

Standard mean-field substitution, with $B^\dag_{S}$ replaced by $ \exp(i\varphi_S)\sqrt{N_S}$, gives the energy of $N=N_2+N_{-2}+N_1+N_{-1}$ excitons as $\mathcal{E'_N}+N(N-1)\Delta /2$ with, after minimization with respect to the phase $\Phi{=}\varphi_2{+}\varphi_{-2}{-}\varphi_1  {-}\varphi_{-1}$,  
\be\label{14}
\mathcal{E'_N}=
\varepsilon_{bd}(N_1{+}N_{-1})
{-}\Delta_o\Big(\sqrt{N_2N_{-2}}{+}
\frac{|\Delta_o|}{\Delta_o}                   
\sqrt{N_1N_{-1}}\Big)^2\,.
\ee

For $N$ excitons created by photon absorption in $S=\pm1$ bright states, $N=N_{1}^{(0)}+N_{-1}^{(0)}$, the number of created electrons and holes with up or down spins are given by $N_{\pm3/2}^{(0)}=N_{\mp1/2}^{(0)}=N_{\pm1}^{(0)}$.  When dark states exist, these carriers split between bright and dark excitons according to $N_{\pm3/2}=N_{\pm1}+N_{\pm2}$ and $N_{\pm1/2}=N_{\mp1}+N_{\pm2}$.

For long-lived carrier spins compared to the condensate lifetime, the numbers of carriers of each species do not change when condensation occurs; so, $N_{\sigma}=N_{\sigma}^{(0)}$ for $\sigma=(\pm1/2,\pm3/2)$. This leads to $N_{2}=N_{-2}\equiv N_d/2\leqslant N_{\pm1}^{(0)}$: indeed, a $S=2$ exciton is created along with a $S=-2$ exciton by carrier exchange between a $S=1$ and a $S=-1$ exciton. This already shows that the exciton condensate always has a bright part when $N_{1}^{(0)}\neq N_{-1}^{(0)}$, and no dark part when $N_{1}^{(0)}$ or $N_{-1}^{(0)}$ is equal to zero. Moreover, the dark part, when it exists, is such that $N_{2}=N_{-2}$, which makes it unpolarized---which cannot be distinguished from linearly-polarized by using mean-field approximation.

Actually, due to spin relaxation during the building-up of the excitonic system, the relevant spin configuration corresponds to $N_{1}^{(0)}=N_{-1}^{(0)}$, which leads to $N_1= N_{-1} \equiv N_b/2$. The $\mathcal{E'_N}$ energy then reads in terms of dark and bright exciton numbers as
\be\label{15}
\mathcal{E'_N}=
\varepsilon_{bd}N_b
{-}\frac{\Delta_o}{4   }\Big(N_d{+}\frac{|\Delta_o|}{\Delta_o}                   
N_b\Big)^2\, .
\ee
For positive $\Delta_o$, as in 3D, or in 2D when the electron and hole quantum wells are close, $\mathcal{E'_N}$ reduces to  $\varepsilon_{bd}N_b- \Delta_o N^2/4$ which is minimum for $N_b=0$ whatever $N$: the condensate is fully dark whatever the exciton density. By contrast, for negative $\Delta_o$, as in 2D when the electron and hole quantum wells are far apart \cite{alloing,vortex}, $\mathcal{E'_N}$, equal to $\varepsilon_{bd}N_b + |\Delta_o| (N-2N_b)^2/4$, is minimum for a number of bright excitons equal to $N_b=(N-N_{th})/2\geqslant 0$, which imposes an exciton number $N$ larger than a threshold value $N_{th}=\varepsilon_{bd}/|\Delta_o|$. The condensate then has a gray character.

It is worth noting that a very short spin relaxation time would release the constraint $N_{2}=N_{-2}$. The lowest-energy state for $\Delta_o$ negative is then obtained for a fully polarized dark condensate, $N_{2}=0$ or $N_{-2}=0$, whatever the exciton density.

We have already tackled the gray character\cite{cc} of exciton condensate in full generality, that is, without knowing the relation (\ref{eq11}) between
scattering amplitudes. By comparing the effective Hamiltonian (\ref{12}) with Eq.~(6) in Ref.\cite{cc}, we are led to set: $\Delta=v_{db}=2v_{dd}=2v_{bb}$ and $\Delta_o=2g_{db}$; the $\Delta_e$ coupling was neglected because of its zero Born value. For long carrier-spin lifetime, a gray condensate was found to appear under a density increase, whatever the $\Delta_o$ sign. However, if we now take into account the relation (\ref{eq11}), we find that the threshold for the appearance of a bright component becomes infinite when $\Delta_o>0$, due to a prefactor cancellation hard to guess, relation (\ref{eq11}) then making the existence of a gray condensate very borderline. Experimental imperfections affecting the energy degeneracy, like crystal strains, could ultimately produce a gray condensate whatever the sample topology. Spin orbit coupling\cite{Can,Kav} might produce the same result. Nevertheless, the very fundamental relation (\ref{eq11}) leads us to conclude that using dipolar excitons was indeed an excellent idea to evidence exciton condensation by optical means\cite{alloing,vortex}. 

\textbf{To conclude}, we have used the coboson many-body formalism to obtain the ground-state energy of two excitons whatever their carrier spins and, from it, to deduce the various scattering amplitudes for dark and bright excitons. This formalism allows us (i) to understand the effects of fermion exchanges, (ii) to obtain the integral equations fulfilled by the scattering amplitudes for brightness-conserving and brightness-changing channels, (iii) to derive a general relation between them, remarkable for its simplicity, $\Delta=\Delta_e+\Delta_o$. Our approach can be used to study other multi-component condensates.

The present work leads us to predict a darkening of the gray condensate when the distance between electrons and holes decreases, excitons with large dipolar momentum being necessary for a bright component to exist in the condensate. We hope that this prediction will stimulate more challenging experiments on exciton condensation which below a critical electron-hole separation, must always occur in a dark state, hence impossible to ``see" through standard photoluminescence experiments.

M.C. acknowledges many fruitful visits to Academia Sinica and NCKU, Taiwan. S.-Y.S. and Y.-C.C. have also benefited from various visits to INSP in Paris. Work supported in part by a S.-Y.S. three-month CNRS position and by the Ministry of Science and Technology, Taiwan under contract MOST 104-2112-M-001.

\textbf{APPENDIX}

Here are the main steps of the novel procedure to get the brightness-conserving and the brightness-changing amplitudes when the carrier spins are $(1/2,-1/2;3/2,-3/2)$. These carriers can form either two bright excitons $(1,-1)$ or two dark excitons $(2,-2)$, coupled by carrier exchanges. This leads us to look for the two-exciton eigenstate as
  \be
 | \Psi_2\ran=\sum_{ij} b_{ij} B^\dag_{i;1} B^\dag_{j;-1}|v\ran+\sum_{ij} d_{ij} B^\dag_{i;2} B^\dag_{j;-2}|v\ran\, .
\ee
 where $|v\ran$ denotes the vacuum state. When used into the Schr\"{o}dinger equation projected over $\lan v|B_{q;-1}B_{p;1}$, we get \be
0\!=\!(E_{pq}-\mathcal{E}_2)b_{pq}+\sum_{ij}\! \Big(\! \xi(_{p\, i}^{ q\, j})b_{ij}-\xi_e^{exch}(_{p\, i}^{ q \, j})d_{ij}\Big)\,
\ee
with a similar equation when projected over $\lan v|B_{q;-2}B_{p;2}$.

By adding and subtracting the two resulting equations and by writing them for $(p,q)=(0,0)$ and $(p,q)\neq (0,0)$, it becomes easy to recover\cite{S} the two coupled equations (\ref{Deltac+d}) by iteration.

\,\,

\,\,

\textbf{Supplemental Materials}

 \textbf{(1) Links between coboson scatterings}
 
We here present some key results of the coboson many-body formalism that are useful for the present work. Details can be found in Refs.[29,30].

(i) The direct-Coulomb scattering shown in Fig.~1(a) follows from two commutators
\bea
\big[H,B^\dag_i\big]_-&=&E_iB^\dag_i+B^\dag_i\,,\\
\big[V^\dag_i,B^\dag_j\big]_-&=&\sum_{pq}\xi(_{p\,i}^{q\,j}) B^\dag_p B^\dag_q\, .
\eea

(ii) The carrier exchange scatterings shown in Figs.~1(c,d) follow from other two commutators
\bea
\big[B^\dag_p,B^\dag_i\big]_-&=&\delta_{pi}-D_{pi}\,,\\
\big[D_{pi},B^\dag_j\big]_-&=&\sum_{q}\big(\lambda_h(_{p\,i}^{q\,j})+\lambda_e(_{p\,i}^{q\,j}) \big)B^\dag_q\, .
\eea
In previous works that do not concern spins,
 we had set $\lambda_h(_{p\,i}^{q\,j})=\lambda_e(_{p\,j}^{q\,i})\equiv \lambda(_{p\,i}^{q\,j})$ for convenience. 

(iii) From the energy-like direct-Coulomb scattering $\xi(_{p\,i}^{q\,j})$ and the dimensionless Pauli scattering $\lambda(_{p\,i}^{q\,j})$, we can construct two exchange-Coulomb scatterings depending on if the carrier exchange takes place before or after the Coulomb process, namely
\bea
\xi^{in}(_{p\,i}^{q\,j})&=&\sum_{uv}\lambda(_{p\,u}^{q\,v})\xi(_{u\,i}^{v\,j})\, ,\\
\xi^{out}(_{p\,i}^{q\,j})&=&\sum_{uv}\xi(_{p\,u}^{q\,v})\lambda(_{u\,i}^{v\,j})=\big[\xi^{in}(_{i\,p}^{j\,q})\big]^*\,.
\eea

(iv) Exchange-Coulomb scatterings are linked by
\be
\xi^{in}(_{p\,i}^{q\,j})-\xi^{out}(_{p\,i}^{q\,j})=(E_{pq}-E_{ij})\lambda(_{p\,i}^{q\,j})\, ,
\ee
that can be checked by calculating $\lan v|B_pB_q HB^\dag_i B^\dag_j |v\ran$ with $H$ acting on the right and on the left.

(v) The above relation is the key to show that the exchange part $\xi^{exch}(_{i\,p}^{j\,q})$ of the kernel scattering $\zeta(_{i\,p}^{j\,q})$ takes two forms, as written in Eq.~(4). By averaging these two forms, the exchange part takes a symmetric form with respect to ``in" states $(i,j)$ and ``out" states $(p,q)$, 
\be
\xi^{exch}(_{i\,p}^{j\,q})=\xi^{in}(_{p\,i}^{q\,j})+\xi^{out}(_{p\,i}^{q\,j})+\left(\frac{E_{pq}+E_{ij}}{2}-E_{00}\right)\lambda(_{p\,i}^{q\,j})\,,
\ee
  which readily fulfills $\big[\xi^{exch}(_{p\,i}^{q\,j})\big]^*=\xi^{exch}(_{i\,p}^{j\,q})$, as required from time-reversal symmetry.

 \textbf{(2) Two-exciton energy} 

We use the coboson many-body formalism to solve the two-exciton Schr\"{o}dinger equation, $(H-\mathcal{E}_2)|\Psi_2\ran=0$ when they do not form a bound molecular state, knowing the full set of single-exciton eigenstates $(H-E_i)B^\dag_{i;s,m}|v\ran=0$ where $|v\ran$ denotes the vacuum state. 
 The $i$ exciton creation operator reads in terms of electron and hole creation operators, ($a^\dag_{\vk,s},b^\dag_{\vk,m}$), as $B^\dag_{i;s,m}=\sum_{\vk_e,\vk_h}a^\dag_{\vk_e,s}b^\dag_{\vk_h,m}\lan \vk_h, \vk_e|i\ran$. Conversely, $a^\dag_{\vk_e,s}b^\dag_{\vk_h,m}=\sum_iB^\dag_{i;s,m}\lan i|\vk_e, \vk_h\ran$.

\textbf{(1)} \textbf{Carrier spins $(s,-s;m,m)$}\

The relation between free-pair operators and exciton operators allows us to look for the two-exciton eigenstates when the two excitons have same hole spin and opposite electron spins, as 
\bea
|\Psi_2\ran=\sum (\cdots) a^\dag_{\vk_e,s}a^\dag_{\vk'_e,-s}b^\dag_{\vk_h,m}b^\dag_{\vk'_h,m}
\nonumber\\
=\sum_{ij} c_{ij} B^\dag_{i;s,m}
B^\dag_{j;-s,m}|v\ran \,\,\,\,\,\,\,\,\,\,\,\,\,\,\,
\eea 
Knowing\cite{MoniqPhysreport,book} that
\be
(H{-}E_{ij})B^\dag_{i;s,m} B^\dag_{j;{-}s,m}|v\ran =\sum_{pq}B^\dag_{p;s,m} B^\dag_{q;{-}s,m}\xi (_{p\, i}^{ q\, j})|v\ran\, ,
\ee
the two-exciton Schr\"{o}dinger equation appears as 
\be
0{=}\sum_{ij}\Big\{ (E_{ij}{-}\mathcal{E}_2)c_{ij}{+}\sum_{pq} \xi (_{i\, p}^{ j\, q})c_{pq}  \Big\} B^\dag_{i;s,m} B^\dag_{j;-s,m}|v\ran\,. \label{8}
\ee

To determine the $c_{ij}$'s, we project the above equation over $\lan v|B_{q;-s,m}B_{p;s,m}$. Knowing\cite{MoniqPhysreport,book} that
\be
\lan v|B_{q;-s,m}B_{p;s,m}B^\dag_{i;s,m} B^\dag_{j;-s,m} |v\ran
=\delta_{pi}\delta_{qj}-\lambda_h(_{p\, i}^{ q \, j})\, ,\label{eq:BBBB}
\ee
we get
\be
0\!=\!(E_{pq}-\mathcal{E}_2)c_{pq}+\sum_{ij}\! \Big(\! \xi(_{p\, i}^{ q\, j})-\xi_h^{in}(_{p\, i}^{ q \, j})- \lambda_h(_{p\, i}^{ q \, j})(E_{ij}-\mathcal{E}_2) \! \Big)c_{ij}\, .\label{eq:cij}
\ee
Since $\mathcal{E}_1=E_0$, we can rewrite the above equation, for $\Delta=\mathcal{E}_2-2\mathcal{E}_1$, as
\be
0=(E_{pq}-E_{00}-\Delta)c_{pq}+\sum_{ij} \Big( \zeta_h(_{p\, i}^{ q\, j})+ \lambda_h(_{p\, i}^{ q \, j})\Delta  \Big)c_{ij}\,, \label{eq:schrodingerDelta}
\ee
with $\zeta_h=\xi-\xi_h^{exch}$, the exchange part, defined in Eq.~(\ref{eq:exch})
 being constructed on $\lambda_h$. Next, we note that ($\zeta, \lambda, \Delta$) all scale as one over the sample volume; so, the product $\Delta\lambda_h$ is negligible in front of $\zeta_h$ and $(E_{pq}-E_{00})$ for $(p,q)\neq(0,0)$.

In a last step, we write Eq.~(\ref{eq:schrodingerDelta}) for $(p,q)=(0,0)$ and for $(p,q)\neq (0,0)$ and we iterate. This readily yields
\be
0=c_{00}\Big(\hat{\zeta_h}(_{0\,0}^{ 0\,0})- \Delta\Big)\, ,\label{12'}
\ee
where $\hat{\zeta_h}$ is solution of the integral equation (\ref{laddereq}) with  $\zeta$ replaced by $\zeta_h$.
Since $c_{00}$ differs from zero because $(B^\dag_0)^2|v\ran$ constitutes the dominant part of the $|\Psi_2\ran$ ground state, we end for this spin configuration with $\Delta=\hat{\zeta_h}(_{0\,0}^{ 0\,0})$.

When the two excitons have the same electron spin and opposite hole spins, $(s,s;m,-m)$, we obtain the same equation (\ref{12}) with $ \hat{\zeta_h}$ replaced by $\hat{ \zeta_e}$.

\textbf{(2)} \textbf{Carrier spins} $(s,s;m,m)$

 We follow the above procedure with the two-exciton ground state written as $|\Psi_2\ran=\sum_{ij} c_{ij} B^\dag_{i;s,m} B^\dag_{j;s,m}|v\ran$ and we obtain the same equation (\ref{8}) with $-s$ replaced by $s$. Difference comes when projecting over $\lan v|B_{q;s,m}B_{p;s,m}$ because the scalar product now has four terms 
 \bea
\lan v|B_{q;s,m}B_{p;s,m}B^\dag_{i;s,m} B^\dag_{j;s,m} |v\ran\hspace{3cm}\nn\\
=\delta_{pi}\delta_{qj}+\delta_{qi}\delta_{pj}-\lambda_h(_{p\, i}^{ q \, j})-\lambda_e(_{p\, i}^{ q \, j})\, .
\eea
The $\delta$ parts of the scalar product lead to a contribution which reads as
 \be
 (E_{pq}-\mathcal{E}_2)(c_{pq}+c_{qp})+\sum_{ij}\! \Big(\! \xi(_{p\, i}^{ q\, j})+ \xi(_{q\, i}^{ p\, j}) \Big)c_{ij}\, ,
  \ee
  while the exchange parts lead to
   \be
- \sum_{ij}\! \Big(\xi_h^{in}(_{p\, i}^{ q\, j})+ \xi_e^{in}(_{q\, i}^{ p\, j})+\big(\lambda_h(_{p\, i}^{ q\, j})+ \lambda_e(_{q\, i}^{ p\, j})\big)\big(E_{ij}-\mathcal{E}_2\big) \Big)c_{ij}\, .
  \ee
  Using Eq. (\ref{sym11}), we then get, for $\bar c_{pq}=(c_{pq}+c_{qp})/2$,
  \be
0=(E_{pq}-E_{00}-\Delta)\bar c_{pq}+\sum_{ij} \frac{\zeta_h(_{p\, i}^{ q\, j})+ \zeta_e(_{p\, i}^{ q\, j})}{2}\bar c_{ij}\label{16}
\ee
 We end up with $\Delta=\hat\zeta_{eh}(_{00}^{ 00})$ where $\hat\zeta_{eh}$ follows from Eq.~(\ref{laddereq}) with  $\zeta$ replaced by $\zeta_{eh}=(\zeta_{e}+\zeta_{h})/2$.

 \textbf{(3)} \textbf{Carrier spins} $(s,-s;m,-m)$\

  With carrier spins $(1/2,-1/2;3/2,-3/2)$, we can form either two bright excitons $(1,-1)$ or two dark excitons $(2,-2)$ which are coupled by carrier exchanges. This leads us to look for  $ |\Psi_2\ran$ as
  \be
 | \Psi_2\ran=\sum_{ij} b_{ij} B^\dag_{i;1} B^\dag_{j;-1}|v\ran+\sum_{ij} d_{ij} B^\dag_{i;2} B^\dag_{j;-2}|v\ran\, .
\ee
 When used into the two-exciton Schr\"{o}dinger equation, we get essentially the same equation (\ref{8}) with a bright part plus a dark part. Difference again comes when projecting this equation over bright and dark exciton states because
 \be
\lan v|B_{q;-S}B_{p;S}B^\dag_{i;S} B^\dag_{j;-S} |v\ran
=\delta_{pi}\delta_{qj}
\ee
for $S=(1,2)$, while 
\be
\lan v|B_{q;-1}B_{p;1}B^\dag_{i;2} B^\dag_{j;-2} |v\ran
=-\lambda_e(_{p\, i}^{ q \, j})\, 
\ee
This gives the projection over $\lan v|B_{q;-1}B_{p;1}$ as
\be
0\!=\!(E_{pq}-\mathcal{E}_2)b_{pq}+\sum_{ij}\! \Big(\! \xi(_{p\, i}^{ q\, j})b_{ij}-\xi_e^{exch}(_{p\, i}^{ q \, j})d_{ij}\Big)\, .
\ee
By symmetrizing this equation with respect to $(p,q)$, we get, for $\bar b_{pq}=(b_{pq}+b_{qp})/2$ and $\bar d_{pq}=(d_{pq}+d_{qp})/2$,
\be
0=(E_{pq}-\mathcal{E}_2)\bar b_{pq}+\sum_{ij} \Big( \xi(_{p\, i}^{ q\, j})\bar b_{ij}-\xi_e^{exch}(_{p\, i}^{ q \, j})\bar d_{ij}\Big)\, ,
\ee
where $\xi_e^{exch}$ can be replaced by $\xi_{eh}^{exch}$ since $\xi_e^{exch}(_{p\, i}^{ q \, j})=\xi_h^{exch}(_{p\, j}^{ q \, i})$. The projection over $\lan v|B_{q;-2}B_{p;2}$ produces a similar equation with ($b,d)$ interchanged. 

By adding and subtracting these two coupled equations, we obtain 
\be
0=(E_{pq}-E_{00}-\Delta)\bar a_{pq}^{(\pm)}+\sum_{ij} \zeta_{eh}^{(\pm)}(_{p\, i}^{ q \, j})\bar a_{ij}^{(\pm)}\, ,
\ee
where $\bar a_{pq}^{(\pm)}=(\bar b_{pq}\pm\bar d_{pq})/2$ and $ \zeta_{eh}^{(\pm)}=\xi\mp\xi_{eh}^{exch}$. This leads to $0=\bar a^{(\pm)}_{00}\Big(\hat\zeta^{(\pm)}_{eh}(_{00}^{ 00})- \Delta\Big)$, where $\hat{\zeta}_{eh}^{(\pm)}$ again follows from Eq.~(\ref{laddereq}) with  $\zeta$ replaced by $\zeta_{eh}^{(\pm)}$.


\begin{thebibliography}{99}

\bibitem{stamper}D. Stamper-Kurn and M. Ueda, Rev. Mod. Phys. {\bf 85}, 1191 (2013).
\bibitem{leggett}A. J. Leggett, {\it Quantum Liquids}, Oxford University Press, Oxford (2006).

\bibitem{cbc}M. Combescot, O. Betbeder-Matibet, and R. Combescot, Phys. Rev. Lett. {\bf 99}, 176403 (2007).
\bibitem{cl}M. Combescot and M. Leuenberger, Solid State Comm. {\bf149}, 567 (2009).
\bibitem{cc}R. Combescot and M. Combescot, Phys. Rev. Lett. {\bf 109}, 026401 (2012).

\bibitem{Shilo_2013} Y. Shilo, K. Cohen, B. Laikhtman, R. Rapaport, K. West, L. Pfeiffer, Nat. Comm. \textbf{4}, 2335 (2013).

\bibitem{alloing}M. Alloing, M. Beian, M. Lewenstein, D. Fuster, Y. Gonzalez, L. Gonzalez, R. Combescot, M. Combescot, and F. Dubin, Europhys. Lett. {\bf 107}, 10012 (2014).

\bibitem{Cohen_2016} K. Cohen, Y. Shilo, K. West, L. Pfeiffer, and  R. Rapaport, Nanoletters \textbf{16}, 3726 (2016).

\bibitem{beian} M. Beian, M. Alloing, R. Anankine, E. Cambril, C. G. Carbonell, A. Lemaitre, and F. Dubin, arXiv:1506.08020.

\bibitem{vortex} R. Anankine, M. Beian, S. Dang, M. Alloing, E. Cambril, K. Merghem, C. G. Carbonell, A. Lemaitre, and F. Dubin, arXiv:1606.04755.

\bibitem{Keldysh1968} L. V. Keldysh and A. N. Kozlov, Sov. Phys. JETP {\bf27}, 521 (1968).
\bibitem{HaugPRB1975} H. Haug and E. Hanamura, Phys. Rev. B {\bf11}, 3317 (1975).
\bibitem{ShumwayPRB2001} J. Shumway and D. M. Ceperley, Phys. Rev. B {\bf63}, 165209 (2001); Solid State Comm. {\bf134}, 19 (2005).
.
\bibitem{Noz} C. Comte and P. Nozi\`{e}res, J. Physique {\bf43}, 1069 (1982).

\bibitem{SCC2016} S.-Y. Shiau, M. Combescot, and Y.-C. Chang, Phys. Rev. A. {\bf 94}, 052706 (2016).


\bibitem{Pieri2000} P. Pieri and G. C. Strinati, Phys. Rev. B {\bf61}, 15370 (2000).
\bibitem{Petrov2004}D. S. Petrov, C. Salomon, and G. V. Shlyapnikov, Phys. Rev. Lett. {\bf93}, 090404 (2003); Phys. Rev. A {\bf 71}, 012708 (2005).

\bibitem{Brodsky2006} I. V. Brodsky, M. Yu. Kagan, A. V. Klaptsov, R. Combescot, and X. Leyronas, Phys. Rev. A {\bf73}, 032724 (2006).
\bibitem{Birse2011} M. C. Birse, B. Krippa, and N. R. Walet, Phys. Rev. A {\bf83}, 023621 (2011).
\bibitem{Alzetto2013} F. Alzetto, R. Combescot, and X. Leyronas, Phys. Rev. A {\bf87}, 022704 (2013).


\bibitem{PlatzmannPRB1994} P. M. Platzmann and A. P. Mill, Jr., Phys. Rev. B {\bf49}, 454 (1994).
\bibitem{Ivanov2001} I. A. Ivanov, J. Mitroy, and K. Varga, Phys. Rev. Lett. {\bf87}, 063201 (2001); Phys. Rev. A {\bf65}, 022704 (2002).
\bibitem{AdhikariPL2002} S. K. Adhikari, Phys. Lett. A {\bf294}, 308 (2002).
\bibitem{Chakraborty2004} S. Chakraborty, A. Basu, and A. Ghosh, Nucl. Instrum. Meth. B {\bf221}, 112 (2004).
\bibitem{Daily2015} K. M. Daily, J. von Stecher, and C. H. Greene, Phys. Rev. A {\bf91}, 012512 (2015).

\bibitem{Cassidy2005} D. B. Cassidy, S. H. M. Deng, R. G. Greaves, T. Maruo, N. Nishiyama, J. B. Snyder, H. K. M. Tanaka, and A. P. Mills, Jr., Phys. Rev. Lett. {\bf95}, 195006 (2005).

\bibitem{Avetissian2014} H. K. Avetissian, A. K. Avetissian, and G. F. Mkrtchian, Phys. Rev. Lett. {\bf 113}, 023904 (2014).

\bibitem{Wang2014} Y.-H. Wang, B. M. Anderson, and C. W. Clark, Phys. Rev. A {\bf89}, 043624 (2014).

\bibitem{MoniqPhysreport} M. Combescot, O. Betbeder-Matibet, and F. Dubin, Physics Reports {\bf 463}, 215 (2008).

\bibitem{book} M. Combescot and S.-Y. Shiau, \textit{Excitons and Cooper Pairs}, Oxford University Press, Oxford (2015).








\bibitem{Fetter} A. L. Fetter and J. D. Walecka, {\it Quantum Theory of Many-Particle Systems} (McGraw-Hill, New York 1971).
\bibitem{S} See Supplemental Materials.

\bibitem{Nous2016}M. Combescot and O. Betbeder-Matibet,	
Europhys. Lett. {\bf58}, 87 (2002).

\bibitem{OBM} O. Betbeder-Matibet and M. Combescot, Eur. Phys. J. B {\bf31}, 517 (2003).
\bibitem{MA}  M. Combescot, M.A. Dupertuis, O. Betbeder-Matibet,
Europhys. Lett. {\bf79}, 17001 (2007)
\bibitem{CCAD} M. Combescot, R. Combescot, M. Alloing, and F. Dubin, Phys. Rev. Lett. {\bf114}, 90401 (2015). 

\bibitem{SYAnnals} S.-Y. Shiau, M. Combescot, and Y.-C. Chang, Annals of Physics {\bf336}, 309 (2013).



\bibitem{Can} M. Ali Can and T. Hakioglu, Phys. Rev. Lett. {\bf 103}, 086404 (2009).

\bibitem{Kav} M. Matuszewski, T. C. H. Liew, Y. G. Rubo, and A. V. Kavokin, Phys. Rev. B {\bf 86}, 115321 (2012).



\end{thebibliography}
\end{document}